\documentclass{elsart}
\usepackage{graphics,amsfonts,amssymb,graphicx}
\begin{document}
\begin{frontmatter}
\title{The quasi-periodic doubling cascade in the transition to weak turbulence}
\author{L. van Veen}
\address{Department of Mathematical and Statistical Science, La Trobe University,\\ Victoria 3086, Australia}
\ead{lvanveen@latrobe.edu.au}

\begin{abstract}
The quasi-periodic doubling cascade is shown to occur in the transition from regular
to weakly turbulent behaviour in simulations of incompressible Navier-Stokes flow on a three-periodic
domain. Special symmetries are imposed on the flow field in order to reduce the computational
effort. Thus we can apply tools from dynamical systems theory such as continuation of
periodic orbits and computation of Lyapunov exponents. We propose a model ODE for
the quasi-period doubling cascade which, in a limit of a perturbation parameter to zero,
avoids resonance related problems. The cascade we observe in the simulations is then
compared to the perturbed case, in which resonances complicate the bifurcation scenario.
In particular, we compare the frequency spectrum and the Lyapunov exponents. The
perturbed model ODE is shown to be in good agreement with the simulations of weak turbulence.
The scaling of the observed cascade is shown to resemble the unperturbed case, 
which is directly related to the well known doubling cascade of periodic orbits.
\end{abstract}

\end{frontmatter}

\section{Introduction}

In the absence of boundaries the incompressible Navier-Stokes equations are symmetric
under rotations, translations and reflections. If we impose a subgroup of symmetries
on the solutions we reduce the number of degrees of freedom in simulations of the flow.
In the early days of numerical simulation such reduction by symmetry was used to probe
into turbulent flow, revealing structures such as the Taylor-Green vortex \cite{brachet}.
Kida \cite{kida1} put forward what is probably the maximal reduction still allowing
for turbulent flow. The corresponding flow is called {\em high symmetric} and was
used to study flow statistics at moderate to high Reynolds number \cite{kida3,kida4,boratav}.

As computers have grown considerably since then, those results may be reproduced today in
simulations without any special symmetry. However, reduction by symmetry can still be
a useful approach when going to ever higher Reynolds number, requiring higher resolution
of the models, and when applying tools of dynamical systems theory to turbulence.

Recently considerable effort has been made to apply continuation and bifurcation analysis
to fluid dynamical problems, see e.g. \cite{lust,sanchez}. If this could be done at realistic
resolution the results could prove a valuable complement to statistical analysis of direct
numerical simulation. At the moment the maximum number of degrees of freedom tackled 
successfully is of order $10^4$. Given that the simulation of turbulent flow at moderate
Reynolds number requires a number of degrees of freedom in the order of $10^6$, reduction
by symmetry can be successfully applied.

Here, we examine the transition from regular to weakly turbulent motion in high symmetric
flow. Exploiting the divergence free condition in addition to the symmetry we gain
a factor of about $300$ with respect to general, non symmetric flow and a factor of
$3/2$ with respect to earlier work on high symmetric flow in terms of the
number of degrees of freedom. Thus we can analyse the transition using continuation of
periodic orbits and computation of Lyapunov exponents, using the direct method for
integration rather then the pseudo-spectral method.

In addition to the Ruelle-Takens scenario, reported on previously \cite{kida2}, we find
that a cascade of quasi-periodic doubling bifurcations, otherwise knows as torus doubling
bifurcations, occurs in the transition to weak turbulence. 
To the author's best knowledge this is the first time this bifurcation sequence is found
in Navier-Stokes flow at realistic truncation level.
In order to relate the numerical
results to theoretical predictions we propose a model Ordinary Differential Equation (ODE) for this
cascade. In contrast to model equations studied previously \cite{kaneko,arneodo} these
equations, based on a fixed time smooth suspension of the H\'enon
map, display a complete cascade with a fixed frequency ratio of the bifurcating tori in 
the limit of a perturbation parameter to zero. If we introduce a slight perturbation in
the model ODE the bifurcation cascade is interrupted, as it is in the simulations of 
weak turbulence. Also, the frequency spectrum at the onset of chaos and the dependence of
the leading Lyapunov exponents on the bifurcation parameter in the simulations is shown to be very 
similar those of the perturbed model ODE.

To within the accuracy of the numerics, the scaling of the cascade is shown to agree
with that of the unperturbed model ODE, which is known theoretically.
When looking at the leading Lyapunov exponents a difference becomes apparent:
in the unperturbed model ODE the largest two nonzero exponents become equal on
an open interval in parameter space in between successive doubling bifurcation points,
whereas in the perturbed model ODE, and also in the flow simulations,
they remain separate. This difference might by caused by the crossing of resonance tongues
or other codimension one phenomena in which the stable torus loses normal hyperbolicity.
A thorough study of the model ODE with
finite perturbation might yield an explanation.

\section{The vorticity equation for high symmetric flow}

\begin{figure}
\begin{center}
\includegraphics[width=0.6\textwidth]{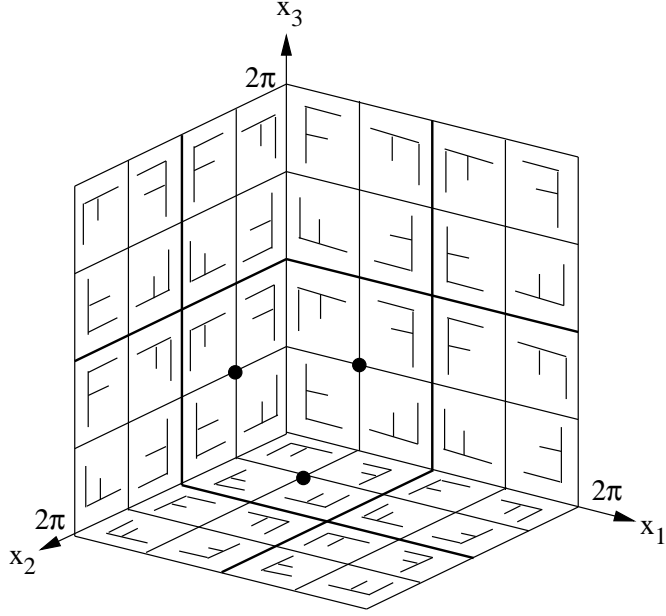}
\end{center}
\caption{The structure of high symmetry. F represents the state on a face of a box with 
edges of length $\pi /2$. Reproduced from \cite{kida1}.}
\label{symmetries}
\end{figure}

Consider an incompressible fluid in a periodic box $0<x_1,x_2,x_3\leq 2\pi$. In terms
of the Fourier representation of velocity and vorticity,
\begin{eqnarray}
\mathbf{v} = \mbox{i} \sum_{\mathbf{k}}\tilde{\mathbf{v}}(\mathbf{k})\mbox{e}^{\mbox{\small i} \mathbf{k}
\cdot\mathbf{x}} & \qquad {\mbox{\boldmath ${\omega}$}} = \sum_{\mathbf{k}}\tilde{\mbox{\boldmath ${\omega}$}}
(\mathbf{k})\mbox{e}^{\mbox{\small i} \mathbf{k}\cdot\mathbf{x}}
\label{fourier}
\end{eqnarray}
we have
\begin{eqnarray}
\frac{\mbox{d}}{\mbox{d} t}\tilde{\omega}_{i}(\mathbf{k}) &=& \epsilon_{ijk}k_{j}k_{l} \,
\widetilde{v_{k}v_{l}} -\nu k^2 \tilde{\omega}_{i}(\mathbf{k}) \\
k_{i}\tilde{u}_{i} &=& 0 \\
\tilde{\omega}_{i}(\mathbf{k})&=& -\epsilon_{ijk}k_{j}\tilde{v}_{k}(\mathbf{k})
\label{basiceq}
\end{eqnarray}
where $\nu$ is the kinematic viscosity, $\epsilon_{i,j,k}$ is the permutation symbol
and the tilde denotes the Fourier transform.
In terms of the standard norm energy and enstrophy are given by
\begin{eqnarray}
E = \frac{1}{2}\|\mathbf{v}\|^2 & \qquad Q = \frac{1}{2}\|\mathbf{\omega}\|^2
\end{eqnarray}
respectively.
Now consider the following discrete symmetry operations: $S_{i}$, reflections in the planes 
$V_{i}$ given by $x_{i}=\pi$ and
$R_{i}$, rotations over $\pi/2$ radiants about the lines $l_{i}: x_{j}=\pi/2 \:\mbox{for $j\neq i$}$.
If the flow is invariant under these operations only one out of three components of vorticity
in a volume 
fraction $1/4^3$ needs to be computed to determine the flow on the periodic domain. Figure
\ref{symmetries} gives an impression of the symmetries. We call a flow invariant under
$S_{i}$ and $R_{i}$, first described by Kida \cite{kida1}, {\em high symmetric}.

Symmetry operations $S_{i}$ and $R_{i}$ introduce linear relations between the Fourier components 
of vorticity.
First of all we have
\begin{equation}
\tilde{\omega}_{1}(k_{1},k_{2},k_{3})=\tilde{\omega}_{2}(k_{3},k_{1},k_{2})=\tilde{\omega}_{3}(k_{2},k_{3},k_{1})
\label{omega1}
\end{equation}
so that we may consider only one component. This scalar function is even or odd 
in its arguments:
\begin{eqnarray}
\tilde{\omega}_{1}(k_{1},k_{2},k_{3}) &=& \tilde{\omega}_{1}(-k_{1},k_{2},k_{3})= \nonumber \\
 & & -\tilde{\omega}_{1}(k_{1},-k_{2},k_{3})=-\tilde{\omega}_{1}(k_{1},k_{2},-k_{3})
\label{omega2}
\end{eqnarray}
and finally we have
\begin{equation}
\tilde{\omega}_{1}(k_{1},k_{2},k_{3})=\left\{ \begin{array}{ll}
-\tilde{\omega}_{1}(k_{1},k_{3},k_{2}) & \mbox{for $k_{1}$ and $k_{2}$ and $k_{3}$ even,} \\
\tilde{\omega}_{1}(k_{1},k_{3},k_{2}) & \mbox{for $k_{1}$ and $k_{2}$ and $k_{3}$ odd,} \\
0 & \mbox{otherwise.} \end{array} \right.
\label{omega3}
\end{equation}
We consider a cubic truncation, i.e. $|k_{1,2,3}|\leq N$. Relations (\ref{omega1}-\ref{omega3})
reduce the number of independent Fourier modes of vorticity by a factor of $192$ in the leading
order, that is $N^3$. This reduction was exploited by Kida \etal \cite{kida3,kida4} to
investigate scaling laws at moderate to high Reynolds number. In the present work we
exploit the divergence free condition for vorticity to further reduce the number of
modes. With the aid of equation (\ref{omega1}) it reads
\begin{equation}
k_{1}\tilde{\omega}_{1}(k_{1},k_{2},k_{3})+k_{2}\tilde{\omega}_{1}(k_{2},k_{3},k_{1})+
k_{3}\tilde{\omega}_{1}(k_{3},k_{1},k_{2})=0
\label{omega4}
\end{equation}
Taking maximal advantage of relations (\ref{omega1}-\ref{omega4}) we consider only Fourier
components of $\omega_{1}$ in the fundamental domain $\{\mathbf{k}\in \mathbb{Z}^3 |
k_{3}>k_{2}, k_{3}\geq k_{1}, k_{1}\geq 0, k_{2}>0, k_{3}\leq N \}$. 
The number of independent modes is reduced by a factor
of $288$ in the leading order.

These components satisfy the following equation
\begin{eqnarray}
\frac{\mbox{d}}{\mbox{d} t}\tilde{\omega}_{1}(k_{1},k_{2},k_{3}) &=& k_{2}k_{3}(\tilde{S}(k_{3},k_{1},k_{2})-\tilde{S}(k_{2},k_{3},k_{1})) \nonumber \\
 & & +k_{1}k_{2}\tilde{T}(k_{2},k_{3},k_{1}) -k_{3}k_{1}\tilde{T}(k_{3},k_{1},k_{2})\nonumber \\
 & & +(k_{2}^{2}-k_{3}^{2})\tilde{T}(k_{1},k_{2},k_{3}) -\nu k^2 \tilde{\omega}_{1}(k_{1},k_{2},k_{3})
\label{domegadt}
\end{eqnarray}
where $\tilde{S}$ and $\tilde{T}$ are the Fourier transforms of
\begin{eqnarray}
S(x_{1},x_{2},x_{3}) &=& v_{1}(x_{1},x_{2},x_{3})^{2} \nonumber \\
T(x_{1},x_{2},x_{3}) &=& v_{1}(x_{2},x_{3},x_{1})v_{1}(x_{3},x_{1},x_{2})
\label{ST}
\end{eqnarray}
and
\begin{equation}
k^2 \tilde{v}_{1}=k_{2}\tilde{\omega}_{1}(k_{3},k_{1},k_{2})-k_{3}\tilde{\omega}_{1}(k_{2},k_{3},k_{1})
\label{vfromomega}
\end{equation}
Energy in supplied by fixing the low order odd mode $\tilde{\omega}_{1}(1,1,3)=-3/8$. Thus we obtain
a family of dynamical systems with one parameter, $\nu$, and a number of degrees of freedom given by
\begin{equation}
n(N)=\left\{\begin{array}{ll}
\frac{2}{3}\left( \frac{N}{2} \right)^3 +\frac{1}{2} \left( \frac{N}{2} \right)^2
-\frac{7}{6} \frac{N}{2}-1 & \mbox{if $N$ is even} \\
\frac{2}{3}\left( \frac{N-1}{2} \right)^3 +\frac{3}{2}\left( \frac{N-1}{2} \right)^2
-\frac{1}{6}\left( \frac{N-1}{2} \right)-1 & \mbox{if $N$ is odd.}
\end{array} \right.
\end{equation}

\section{Numerical considerations}

In performing time integrations we avoid the use of a pseudo-spectral method, commonly employed
for three dimensional simulations of Navier-Stokes. Up to a truncation level of $N=25$ ($n=1365$)
the direct method is not exceedingly slow due to the reduction of the number of degrees 
of freedom described above and yields easy access to the
Jacobian for integration of the variational equations. 
Also, the direct integration code can easily be run in parallel, distributing the computation of
the components of the vector field over any number of processors.
For higher truncation levels, at which storage of the nonlinear interaction coefficients
requires huge memory space, it is mandatory
to use a pseudo spectral method, as described in \cite{kida3},
even though it does not exploit the divergence free condition
(\ref{omega4}). For time integration a seventh to eight order Runge-Kutta-Felbergh scheme with
adaptive step size is employed. 
As many of the results presented here
are based on rather long time integrations it important to keep the error tolerance low.
We have checked energy conservation at zero forcing and viscosity using the high order
and the fourth order Runge-Kutta schemes. For realistic $\mbox{O}(1)$ levels of the energy,
an integration time $\Delta t= 10^3$ and a fixed error tolerance $\delta E=10^{-9}$
the step size required for the fourth order scheme is about ten times smaller then the
average step size using the high order scheme. As the high order method needs thirteen
evaluations of the vector field at each time step, against four for the low order scheme, 
it is more efficient.

\begin{figure}[t]
\begin{center}
\includegraphics[width=0.9\textwidth]{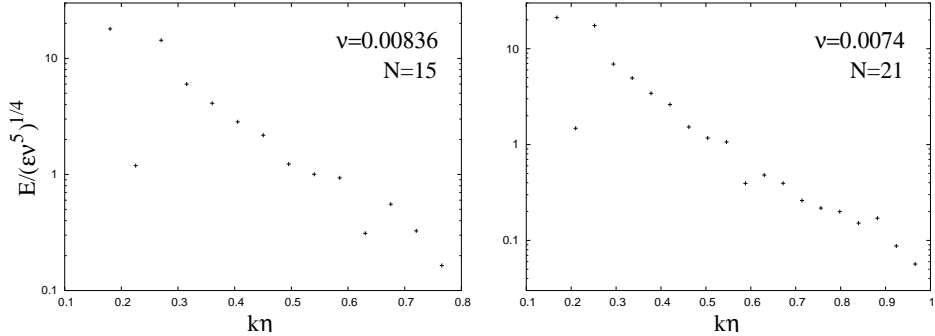}
\end{center}
\caption{Band-averaged energy spectra near the first transitions to chaos, described in section \ref{qpdc}.
Obtained from an integration of $300$ time units, log-linear scale in normalised units. Left:
truncation level $N=15$, right: $N=21$.}
\label{bandav}
\end{figure}

Below simulations are performed with a viscosity in the range $0.005 < \nu < 0.01$ and the 
truncation level fixed to $N=15$ ($n=300$).  We 
computed the energy and the enstrophy, as well as Taylor's micro-scale Reynolds
number, $R_{\lambda}$, and Kolmogorov's dissipation length scale, $\eta$, defined by
\begin{eqnarray}
R_{\lambda} = \sqrt{\frac{10}{3}}\frac{E}{\nu \sqrt{Q}}
 & \qquad \eta = \sqrt[4]{\frac{\nu^2}{2Q}}
\end{eqnarray}
The dimensionless number $\eta N$ indicates if the resolution is high enough in numerical
simulations. If $\eta N \sim 1$  the truncation error is considered negligible. 
At the fist transition to chaos, the focus of this paper, we have $\eta N\approx 0.8$.
The band average energy spectrum is shown in figure \ref{bandav}(left). 
In order to make sure that the results presented below do not depend critically on the truncation
level we repeated some of the computations, in particular the location of the quasi-periodic
bifurcation points described in section \ref{qpdc}, at $N=21$ ($n=814$). 
The qualitative behaviour remains the same
as the bifurcation points shift slightly to lower viscosity. For this truncation level we
have $\eta N\approx 1$ at the first transition point. The band-averaged
energy spectrum is shown in figure \ref{bandav}(right). At the small
scales  an exponential decay is visible, indicating that our numerical results are reliable.

Periodic orbits are continued in the viscosity as fixed points of a Poincar\'e map, using 
the arclength method
as described in \cite{simo}. This method is time consuming because of the integration of the
linearised equations. However, running the code in parallel this is no obstacle.

\section{Transitions to chaos}
\label{transitions}
\begin{figure}[p]
\begin{center}
\includegraphics[width=\textwidth]{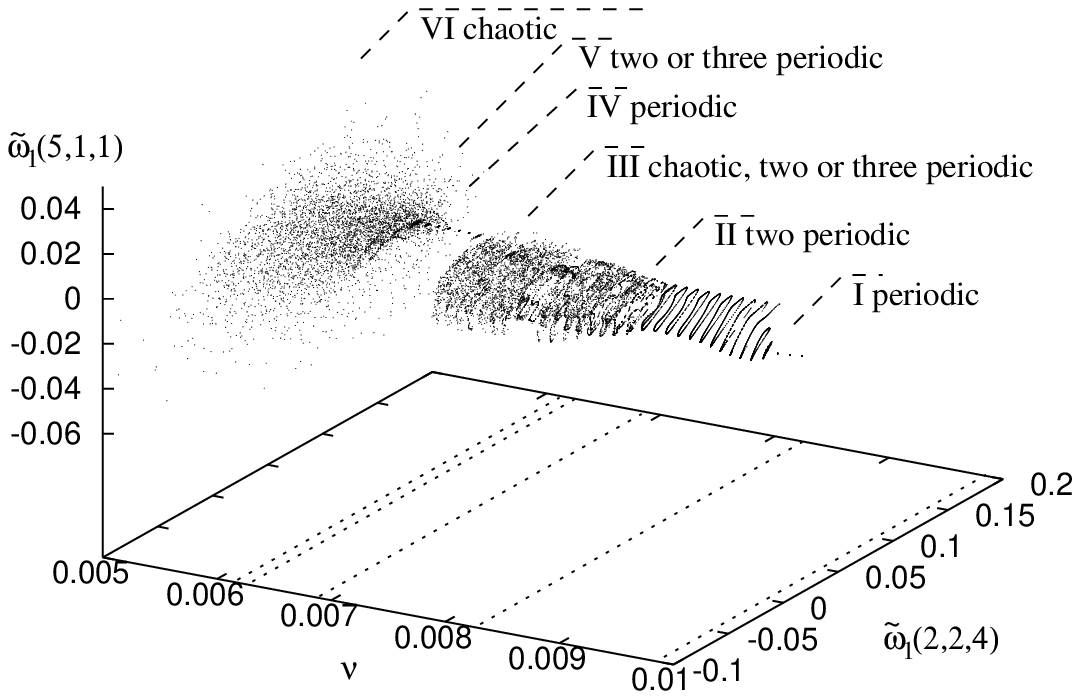}
\end{center}
\caption{Limit point diagram of high symmetric flow simulations. Iterates of the Poincar\'e map,
obtained from an integration of length $\Delta t=1000$, are drawn for each parameter value. 
Transient motion is discarded and the step size of the parameter is set to $\delta \nu=1\cdot 10^{-4}$.
The last point of each integration is the initial point at the next parameter value, starting at 
the stable periodic orbit at $\nu=0.0105$. Visible is the transition I$\rightarrow$II
from periodic to quasi periodic, II$\rightarrow$III to chaotic, III$\rightarrow$IV back to periodic,
IV$\rightarrow$V to quasi periodic and V$\rightarrow$VI to chaotic/turbulent. In region III the
behaviour alternates between chaotic and two or three periodic while the spatial structure of the 
flow remains simple.}
\label{trans}
\end{figure}
\begin{figure}[p]
\begin{center}
\includegraphics[width=0.6\textwidth]{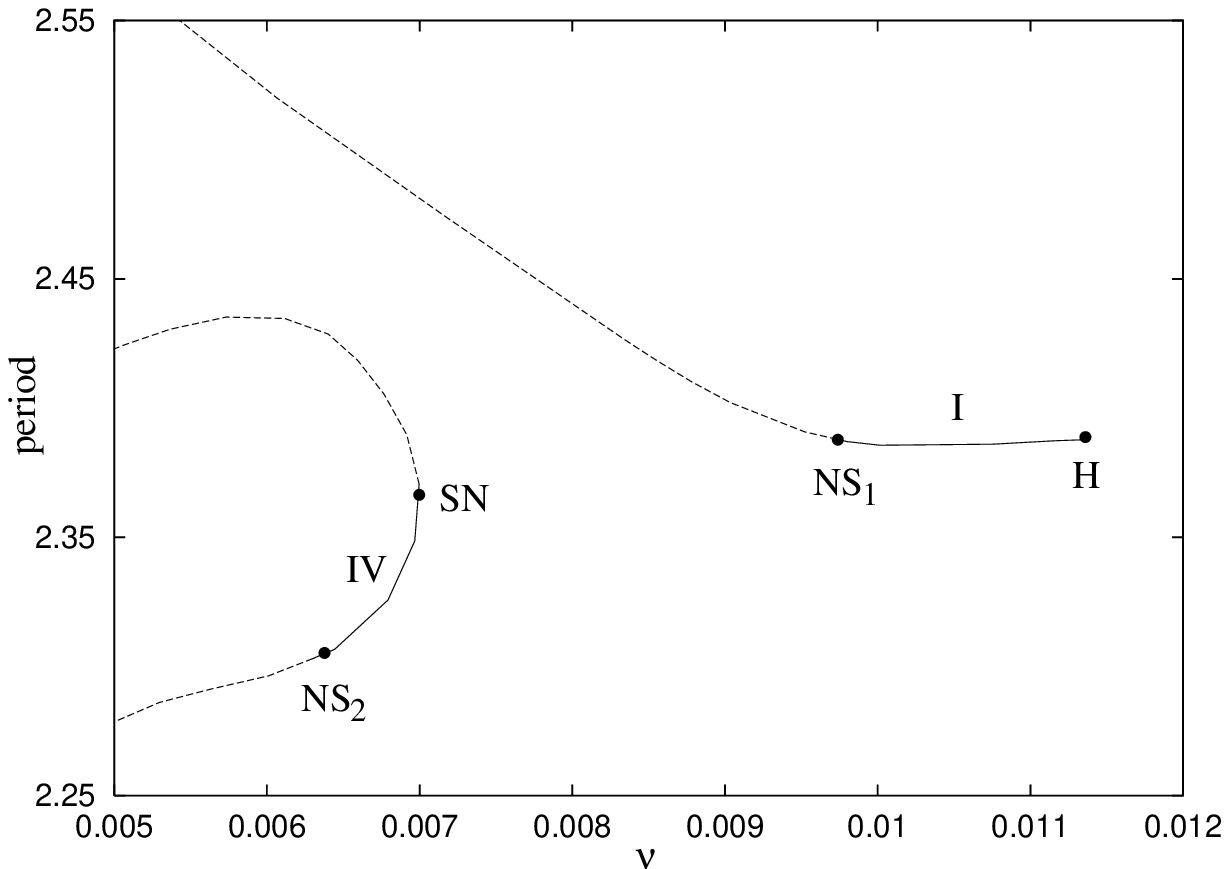}
\end{center}
\caption{Branches of periodic orbits of high symmetric flow, truncated at $N=15$. Solid lines denote stable
branches and dashed lines unstable branches. Bifurcation points are marked SN for saddle-node, 
NS for Neimark-Sacker and H for Hopf. The roman number refer to the stable periodic behaviour as seen in
the limit point diagram, figure \ref{trans}.}
\label{branches}
\end{figure}

For large viscosity an equilibrium state is the global attractor in our simulations of high symmetric flow.
At $\nu \approx 0.0113$
a Hopf bifurcation occurs in which a stable periodic orbit is created. To get an impression of the
transitions from periodic to chaotic motion we computed a limit point diagram  in the range
$0.01 > \nu > 0.005$ of the Poincar\'e map on the coordinate plane given by $\tilde{\omega}(0,2,4)=-0.05$. 
In this range the time average micro-scale 
Reynolds
number varies from $R_{\lambda}\approx 50$ to $R_{\lambda}\approx 27$, indicating that the
aperiodic behaviour can be classified as weak turbulence, fully developed turbulence sets
in around $\nu\approx 0.004$ and $R_{\lambda}\approx 60$ in high symmetric flow.
The limit point diagram is shown in figure \ref{trans}.
Two parameter ranges with periodic behaviour can be seen: one around $\nu = 0.01$ and one around
$\nu = 0.0068$. A continuation of these periodic orbits in parameter $\nu$ is shown in figure \ref{branches}.
The branch which is stable around $\nu = 0.0068$ does not bifurcate from an equilibrium at high viscosity.
Both branches become unstable in a Neimark-Sacker bifurcation. Directly beyond 
these bifurcation points we expect the behaviour to be quasi periodic, and indeed invariant circles
appear in the Poincar\'e section, figure \ref{trans}. The breakdown of these invariant tori gives
rise to chaos. The torus created at $\mbox{NS}_{2}$ displays a quasi-periodic Hopf bifurcation and
thus the Ruelle-Takens scenario to chaos is followed, as reported in \cite{kida2}. The torus created 
at $\mbox{NS}_{1}$ displays a quasi-periodic doubling bifurcation which turns out to be the first
of a cascade, discussed in detail in section \ref{qpdc}.

\begin{figure}
\begin{center}
\includegraphics[width=0.8\textwidth]{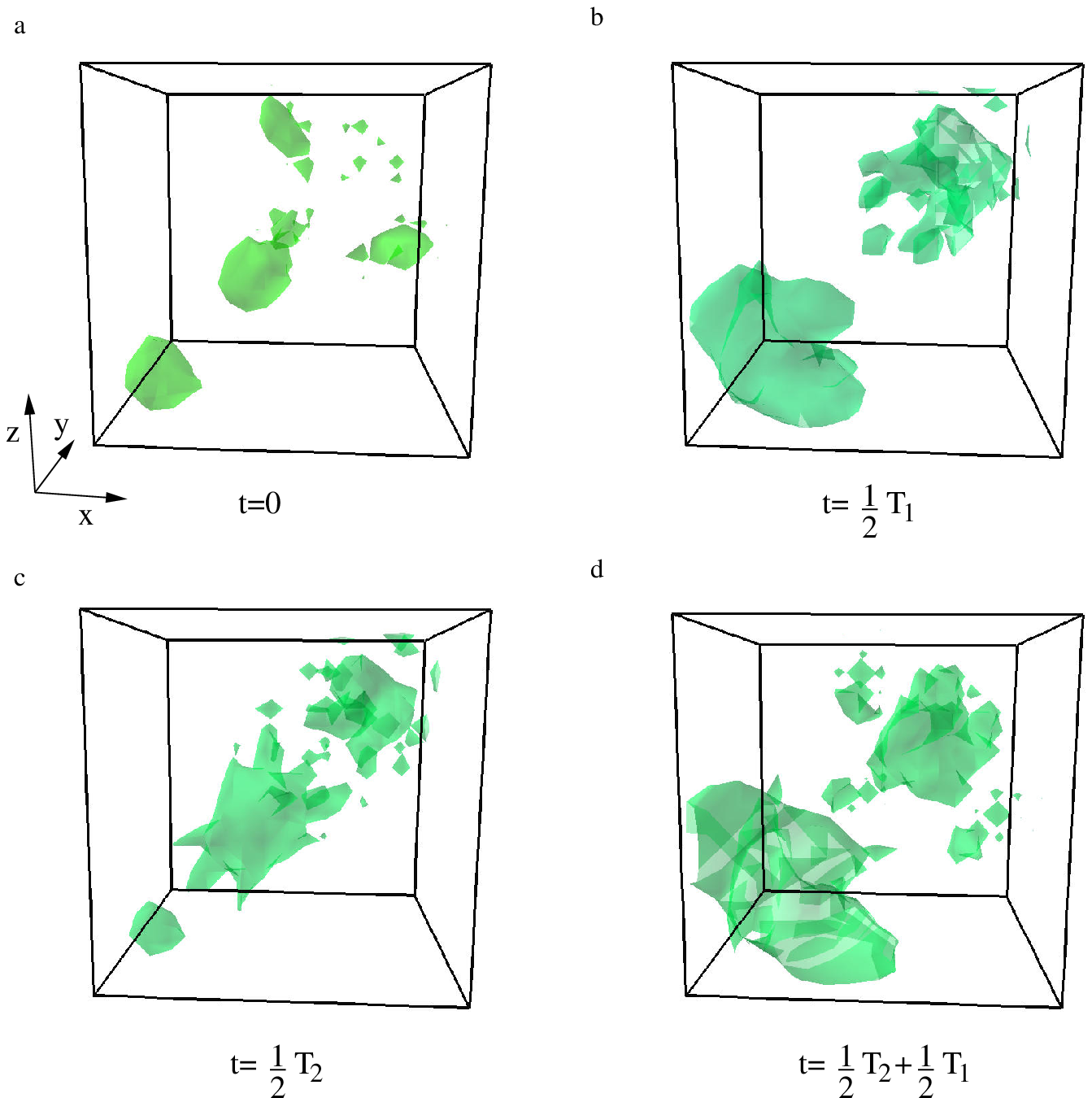}
\end{center}
\caption{Isosurface of enstrophy during quasi-periodic behaviour corresponding to figure \ref{doubling}a.
Edges of length $\pi$, taken from the $2\pi$-periodic domain and isosurface level set to eight times the 
spatio-temporal average. The main diagonal is an axis of three-fold rotational symmetry. The time step
between a and b (c and d) is half the short period $T_{1}=2\pi/\omega_{1}$ and from a to c (b to d)
half the long period $T_{2}=2\pi/\omega_{2}$.}
\label{snapshots}
\end{figure}

The fundamental frequencies of the tori created at $\mbox{NS}_{1,2}$ have a physical interpretation.
The higher frequency, $\omega_{1}\approx 2.6$ is set by the large eddy overturning time,
estimated as $T=l/U\approx 2.3$ where $U\approx 0.9$ is the
root mean square velocity and $l=\sqrt[3]{\pi^3/3}$ is the characteristic length
scale given that by symmetries $S_{i}$, the planes $x_{1,2,3}=\pi$ are impermeable and by
symmetries $R_{i}$
the velocity field has a three fold rotational symmetry around the main diagonal $x_{1}=x_{2}=x_{3}$.
The lower frequency, $\omega_{2}\approx 0.26$, corresponds to a modulation of the amplitude
of the energy. It is also present in fully developed turbulence with high symmetry and can be
though of as a retaining time scale of anomalously high energy. The dynamics on these time scales
is illustrated by the isosurface of enstrophy shown in figure \ref{snapshots}. On the short
time scale, $T$, patches of high vorticity are generated on the main diagonal and off the
diagonal in triples related by the three fold rotational symmetry. These patches converge to the origin 
and to the center of the $2\pi$-periodic box, points of special symmetry. On the longer time
scale the overall amplitude of this process is modulated. Note, that the patches of high vorticity
do not have the elongated, tubular shape typical of developed turbulence.

\section{The quasi-periodic doubling cascade}
\label{qpdc}

\begin{figure}
\begin{center}
\includegraphics[width=\textwidth]{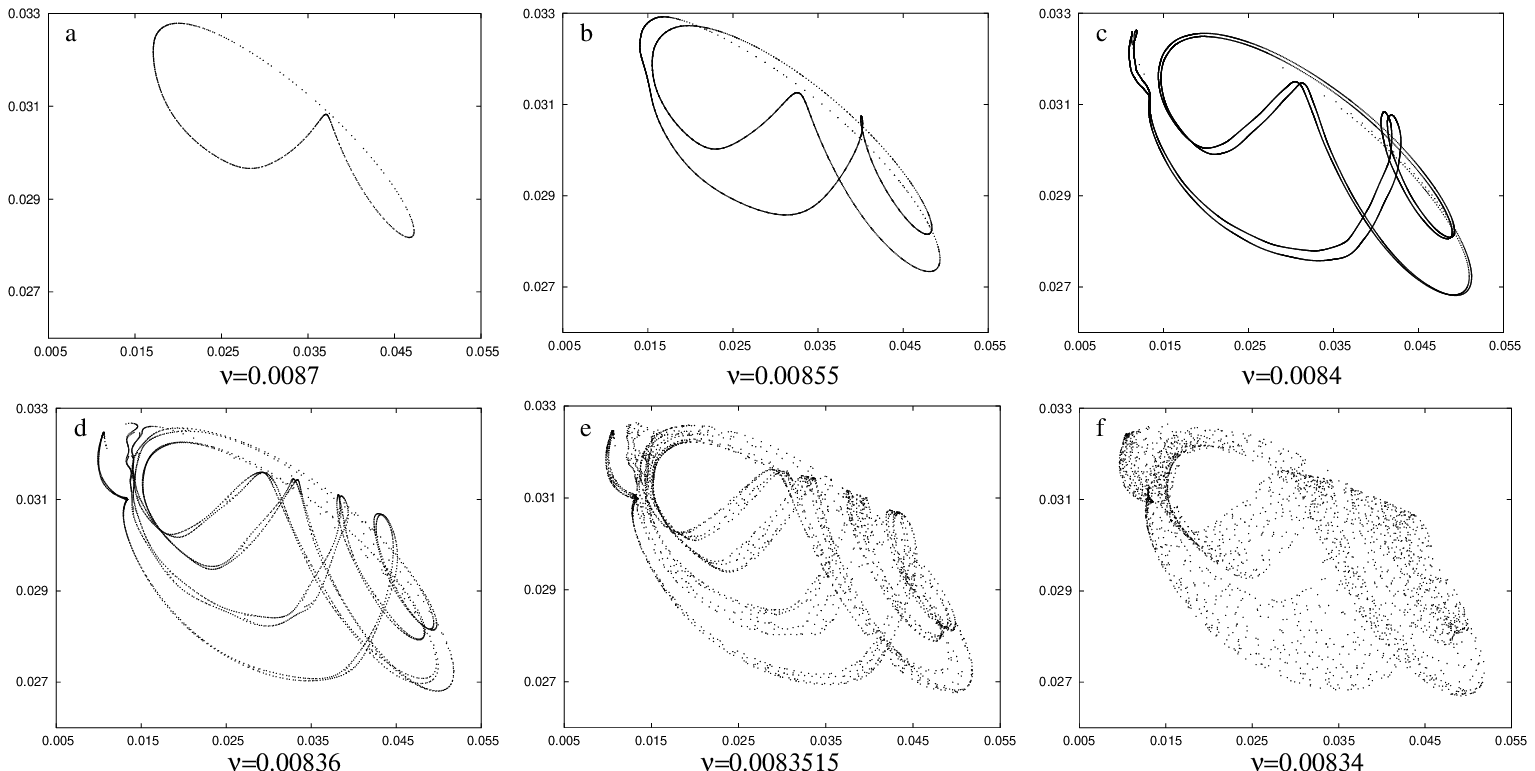}
\end{center}
\caption{Poincar\'e sections in the plane $\omega(0,2,4)=-0.05$, projected onto the modes $\omega(2,2,4)$
and $\omega(3,1,3)$. Four quasi-periodic doublings are shown in (a-e) and the resulting chaotic attractor
in (f).}
\label{doubling}
\end{figure}

In figure \ref{doubling} six Poincar\'e plots are shown beyond bifurcation point $\mbox{NS}_{1}$.
The torus ``doubles'' at least four times (a-e) before a chaotic attractor appears with a
shape very similar to the doubled torus.

Such sequences of torus doubling bifurcations have been observed both numerically, e.g. in 
severe truncations of the Navier-Stokes equations \cite{franceschini} and in 
a periodically driven low-order atmosphere model \cite{broer2}, and experimentally,
e.g. in electronic circuits \cite{anishenko,zhong} and have also been studied in connection 
to strange non chaotic attractors \cite{stag}.
In these
examples chaotic behaviour is observed after two or three doublings.
In our system we find four doublings, after which quasi-periodic motion is very hard
to distinguish from aperiodic motion numerically.

The observation of incomplete quasi-periodic doubling cascades led to the introduction of simple models
which highlight the similarity to the well known periodic doubling cascade \cite{kaneko,arneodo}.
In these models an infinite cascade can only be observed in a limit of the parameters were
the vector field can be decomposed into a three dimensional vector field which exhibits a period
doubling cascade and a constant rotation. Away from this limit the stable torus is more fragile
after each bifurcation and the cascade is interrupted.

In order to see why the cascade is interrupted, we need to investigate
the quasi-periodic doubling bifurcation in detail.
Rigorous bifurcation theorems for bifurcating
invariant tori were formulated by Broer {\sl et al} \cite{broer,braaksma}. They include
the bifurcation shown here, in which a torus loses stability and a new stable torus is
created with one fundamental frequency half that of the original torus. However, the bifurcation
theorem is given for a one parameter family of tori that satisfy a non resonance condition
on the fundamental frequencies. If the fundamental frequencies of the torus become resonant
normal hyperbolicity is lost and persistence under parameter variations is not guaranteed.
We have only one parameter in our system and the fundamental frequencies depend on it
continuously. As we vary the parameter
we cut through a Cantor set of values at which the torus is normally hyperbolic.
Thus, the loss of normal hyperbolicity through resonance is a possible mechanism for
the interruption of the cascade.

However, the resonances we pass through, i.e. the Arnold' tongues we cross, are of fairly
high order as we find for the fundamental frequencies at $\mbox{NS}_{1}$
that $\omega_{1}\approx 10 \omega_{2}$.
Recently an algorithm for the continuation of invariant tori was developed that ``steps over''
such high order resonances without a problem, and in fact this algorithm was tested on
a system that exhibits a cascade of quasi-periodic doublings \cite{schilder}. In this
paper it is pointed out that in the absence of a non resonance condition the notion of a 
quasi-periodic bifurcation point itself is unclear, as no normally hyperbolic torus exists in a whole 
interval in parameter space. On one side of this interval we observe the stable single
torus and on the other side the stable doubled torus. This interval turns out to be small compared to
the numerical resolution of our simulations, a step size in the viscosity of about 
$\delta \nu\approx 10^{-7}$.

Apart form resonances there are other mechanism that can interrupt the cascade.
Stagliano {\sl et al.} \cite{stag}, for instance, concluded that the cascade was interrupted when the stable torus loses smoothness
and collides with an unstable parent torus.
In reference \cite{broer2} a strikingly complicated, yet incomplete picture is sketched of
the mechanisms that can interrupt the cascade.

\begin{figure}
\begin{center}
\includegraphics[width=0.9\textwidth]{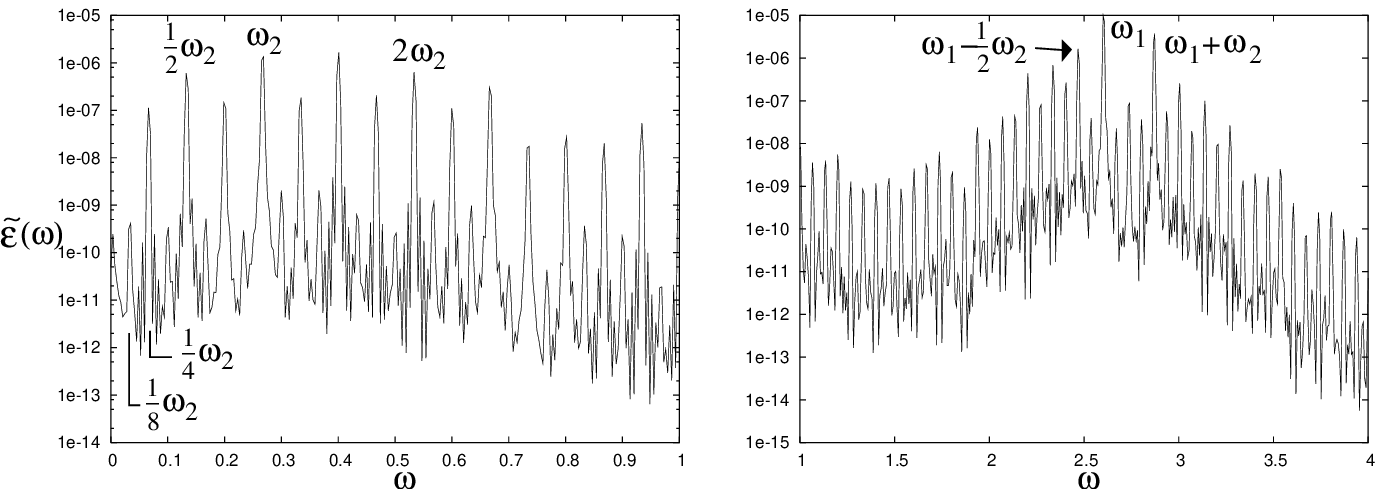}
\end{center}
\caption{Frequency spectrum of the energy just beyond bifurcation point $\mbox{D}_{3}$ around
the widely separated frequencies $\omega_{1,2}$. Note the peaks due to the doubling bifurcations
around $\omega_{2}$ and the resonance peaks around $\omega_{1}$.}
\label{freqspec}
\end{figure}

In order to interpret our numerical results, let us formulate a new model ODE for the
quasi-periodic doubling cascade. Models studied before \cite{kaneko,stag} were
based on the logistic map and cannot be regarded as the Poincar\'e map of a smooth flow.
Our starting point is the H\'enon map.

Let $\mathcal{H}_{s}(x,y)$ be a family of H\'enon maps on $\mathbb{R}^2$
which exhibits a period doubling cascade. Let $\bar{s}_{i}$ be the period doubling bifurcation point 
at which a stable $2^i$-periodic point is created and let $\lim_{i\rightarrow\infty}\bar{s}_{i}=s_{c}$.
Let $\phi_{t}(x,y,u;s)$ be a smooth fixed time suspension of $\mathcal{H}_{s}$ on $\mathbb{R}^2 \times S$ so
that $\phi_{1}(x,y,0;s)=(\mathcal{H}_{s}(x,y),0)$. The flow $\phi_{t}$ is generated by an autonomous
vector field, denoted here by $f_{1}\partial_{x}+f_{2}\partial_{y}+\partial_{u}$.
In appendix \ref{suspension} explicit expressions are given as formulated in \cite{mayer}.
Finally, we add a second periodic variable and a coupling proportional to $\epsilon$:
\begin{eqnarray}
\dot{x} &= f_{1}(x,y,u;s)+\epsilon h_{1}(v)  \qquad \dot{u} &= 1 \nonumber \\  
\dot{y} &= f_{2}(x,y,u;s)+\epsilon h_{2}(v)  \qquad \dot{v} &= \omega +\epsilon g(x,y)
\label{normform}
\end{eqnarray}
with $\mathbf{h}: S\rightarrow \mathbb{R}^2$, $g: \mathbb{R}^2\rightarrow \mathbb{R}$ and $\omega$ irrational. 
For $\epsilon=0$ the flow on $\mathbb{R}^2 \times S \times S$
is given by $\psi_{t}(x,y,u,v;s)=(\phi_{t}(x,y,u;s),v_{0}+\omega t)$ and we know that
\begin{enumerate}
\item At $s=\bar{s}_{i}$ a stable torus with fundamental frequencies $1/2^i$ and $\omega$
is created.
\item The quasi-periodic doubling bifurcations accumulate on zero with the same scaling as
for the period doubling cascade, thus for the bifurcation points we find that $\lim_{i\rightarrow\infty}(s_{i}-s_{i-1})/(s_{i+1}-s_{i})=\delta$, 
where $\delta=4.669\ldots$, the Feigenbaum constant.
\item The Lyapunov exponents of the attracting torus are determined by the Floquet multipliers
of the corresponding periodic points of $\mathcal{H}_{s}$. Therefore the leading nonzero exponent should be equal
to the second exponent on a open interval contained in each interval $(\bar{s}_{i},\bar{s}_{i+1})$.
\end{enumerate}
For finite $\epsilon$ the second fundamental frequency will depend on $s$ continuously and
resonance points will be passed through on the way to the accumulation point $s=s_c$.

In the following we compare the model ODE to the simulations of weak turbulence. The damping
parameter of the H\'enon map is fixed to $j=1/4$. The unperturbed model then displays
a doubling cascade with $s_{1}=1.1718\ldots$, $s_{2}=1.7031\ldots$, $s_{3}=1.8144\ldots$ 
and $s_{c}=1.84525\ldots$. In the perturbed system we fix the second frequency to the golden
mean, i.e. $\omega=(\sqrt{5}+1)/2$ and
\begin{eqnarray}
h_{1}(v) &=& \sin 2\pi v \nonumber \\ 
h_{2}(v) &=& 0 \nonumber \\
g(x,y) &=& y
\end{eqnarray}
The perturbation parameter is fixed to $\epsilon=0.001$. At this perturbation strength we observe
four doubling bifurcations before chaos sets in. In the Poincar\'e intersection plane given by $v=0$,
projected on the $x$ and $y$ variables,
the bifurcation sequence looks very similar to the one shown in figure \ref{doubling}.

We have computed the frequency spectrum near the onset of chaos and the (partial) Lyapunov spectrum
both for the weak turbulence and for the model ODE. 
For computation of Lyapunov exponents of the high-symmetric flow 
we used finite differencing rather than integration 
of the full linearised
system. Integration was done until the zero exponent associated with the direction transversal
to the flow and tangent to the torus had converged up to $10^{-4}$.
\begin{figure}
\begin{center}
\includegraphics[width=\textwidth]{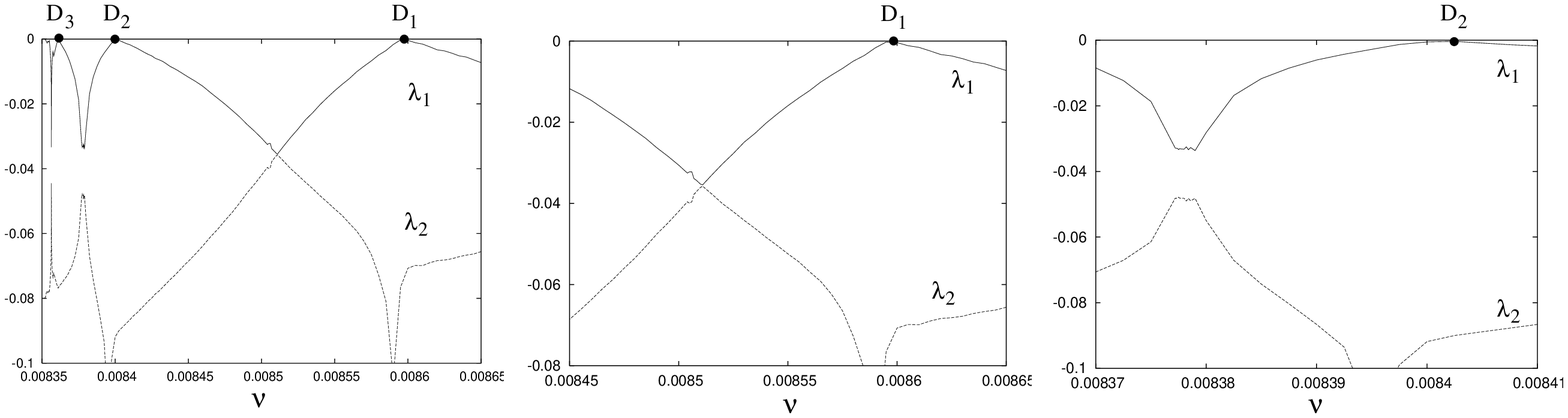}
\end{center}
\caption{Left: largest nonzero Lyapunov exponents $\lambda_{1,2}$ as a function of $\nu$ in the quasi-periodic doubling cascade.
The first doubling points have been marked $\mbox{D}_{i}$, $i=1,2,3$. Right: enlargement of
the regions around $\mbox{D}_{1}$ and $\mbox{D}_{2}$. Note that after the second
doubling the Lyapunov exponents do not meet.}
\label{exponents}
\end{figure}
\begin{figure}
\begin{center}
\includegraphics[width=\textwidth]{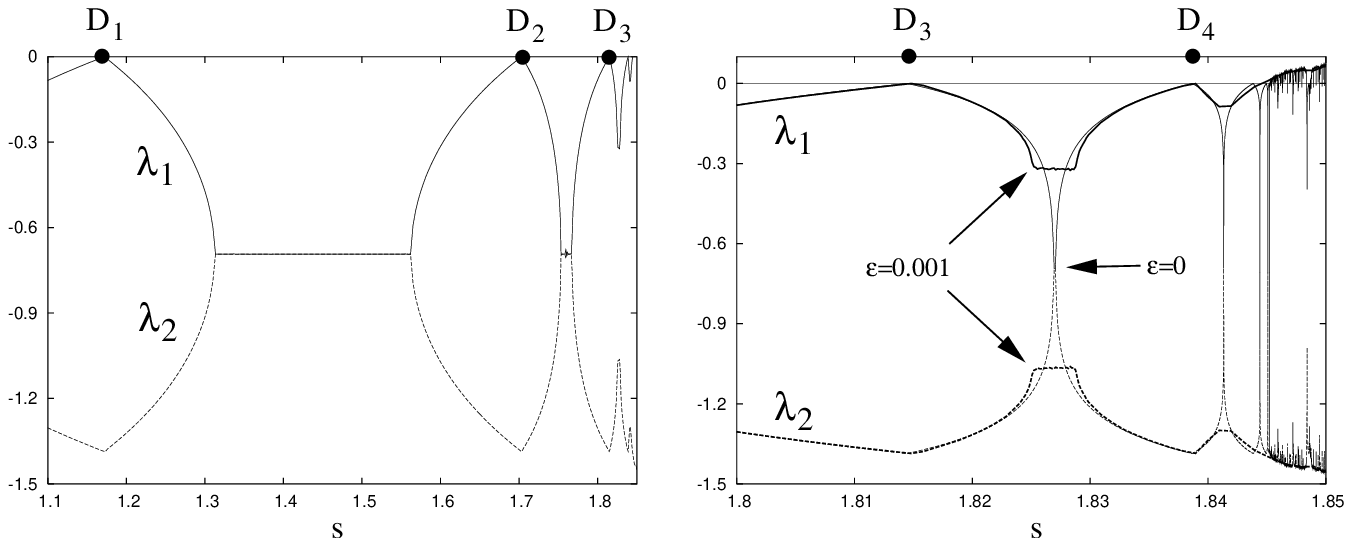}
\end{center}
\caption{Left: largest nonzero Lyapunov exponents $\lambda_{1,2}$ as a function of $s$ for
the model ODE (\ref{normform}) with $\epsilon=0.001$.
The doubling points have been marked as in figure \ref{exponents}. Right: enlargement of
the region around $\mbox{D}_{3,4}$. The bold lines denote the exponents for the perturbed
model at $\epsilon=0.001$. For comparison the exponents for the unperturbed case 
($\epsilon=0$) are shown with thin lines. Note that the exponents of the perturbed system do no meet between
the doubling points.}
\label{susexponents}
\end{figure}

Turning first to the frequency spectrum we observe that after three doublings bifurcations peaks at
frequencies $\omega_{2}/2$, $\omega_{2}/4$ and $\omega_{2}/8$ show up, as well as combination
peaks around $\omega_{1}$. Figure \ref{freqspec} shows the spectrum around the fundamental frequencies
for the simulations of weak turbulence. A rather similar figure for the model ODE is omitted here.
In the case of a period doubling cascade the amplitude ratio $\mu$ between
consecutive peaks $\omega/2^i$ can be predicted on basis of the scaling constants \cite{feig}.
We attempted to measure this ratio for the spectrum shown in figure \ref{freqspec} but only the first
three peaks can be distilled from these data. These scale roughly by a factor of $\mu=4$, not
far from the theoretical prediction for the period doubling case $\mu=4.648\ldots$. The same
estimate is obtained for the model ODE. In the
case of Poincar\'e maps, rather then discrete maps, similar deviations from the theory are found.

In figures \ref{exponents}(left) and \ref{susexponents}(left) the leading Lyapunov exponents are shown as a 
function of the bifurcation parameter, i.e. the viscosity $\nu$ for the high-symmetric flow and
the parameter $s$ of the H\'enon map for the model ODE.
From these
data we can estimate the locus of the bifurcation points, stressing once more that this notion
is ill-defined on a scale smaller then our resolution in parameter space. Thus we find
consecutive estimates for the scaling constant $\delta$ as 
$\delta_{i} = (\bar{\nu}_{i}-\bar{\nu}_{i-1})/(\bar{\nu}_{i+1}-\bar{\nu}_{i})$ or
$\delta_{i} = (\bar{s}_{i}-\bar{s}_{i-1})/(\bar{s}_{i+1}-\bar{s}_{i})$:
\begin{eqnarray*}
\delta_{1} &=& 4.64 \pm 0.2  \ \qquad\delta_{2} = 4.88 \pm 0.5 \qquad \mbox{for high-symmetric flow and}\\
\delta_{1} &=& 4.75 \pm 0.05  \qquad\delta_{2} = 4.63 \pm 0.2 \qquad \mbox{for the model ODE.}
\end{eqnarray*}
Again, this is compatible with the theory for period doubling cascades, albeit a rough estimate.

In figure \ref{exponents}(right) enlargements around the points $\mbox{D}_{1}$ and $\mbox{D}_{2}$
are shown for the high-symmetric flow simulations.
After the first doubling the leading non zero exponents are equal on small interval of
order $\delta \nu=10^{-6}$. After the
second doubling the leading exponents neither cross nor become equal.
This result was confirmed by integration of the full linearised equations at several points
in the parameter range where the leading exponents attain their minimal distance.
The parameter range between the third and fourth doublings is too small to see whether
the exponents meet. In this range 
we record the passage through a high order resonance which does not destroy the torus. 

In figure \ref{susexponents}(right) an enlargement is shown of the Lyapunov exponents of
the model ODE around the doubling points $\mbox{D}_{3}$ and $\mbox{D}_{4}$. For comparison,
the exponents of the unperturbed model, computed from the H\'enon map itself, 
have been drawn with thin lines. Again the Lyapunov exponents do not become equal in between
doubling points. The step size has been taken as small as $\delta s=1\cdot 10^{-5}$.
Apparently, the slightly perturbed model ODE yields the same behaviour of the leading Lyapunov
exponents as the simulations of high-symmetric flow.

For now we have no explanation for the absence of crossing or equal leading Lyapunov
exponents in between doubling points. As noted before,
the invariant tori are structurally stable only on a fractal
domain of the parameter and bifurcations might go unnoticed when monitoring Lyapunov
exponents with finite step size for the parameter. A more thorough understanding could
be obtained by a continuation of the invariant torus itself and by further study of the 
model ODE (\ref{normform}), which is the topic of future research.

\section{Conclusion}

We have investigated the transition from stationary to disordered behaviour in simulations of high symmetric 
flow. In combination with the symmetry the divergence free
condition was exploited to reduce the number of degrees of freedom by
a factor of about $300$ with respect to simulations of general periodic Navier-Stokes flow, a gain of 
$30 \%$ compared to earlier work \cite{kida1}-\cite{boratav},\cite{kida2}.
This allowed us to investigate the bifurcation scenario in detail and at realistic truncation level.

Along with the Ruelle-Takens scenario, the quasi-periodic doubling cascade occurs 
as a route to weak turbulence. By means of Poincar\'e sections, power spectra and Lyapunov
exponents we have shown that this cascade bears close resemblance to the well known doubling
cascade for periodic orbits. We observe four doublings after which quasi-periodic behaviour
is very hard to distinguish from chaotic behaviour numerically. In previous work on the
quasi-periodic doubling cascade, both numerical and experimental, chaos is reported to
set in after two or three doublings \cite{anishenko,franceschini,zhong}. 
The fact that we see more doublings might be due
to the widely separate fundamental frequencies of the bifurcating tori in our system, as
is the case in recent work by Schilder {\sl et al}\cite{schilder}. 

In order to make the notion of a quasi-periodic doubling cascade precise we have proposed
a model ODE, equations (\ref{normform}), such that in the limit of a perturbation 
parameter $\epsilon \downarrow 0$
the correspondence to the periodic doubling cascade is exact. The scaling of the first steps
of the observed quasi-periodic doubling cascade agrees with the prediction of the unperturbed
equations (\ref{normform}). Looking at the leading nonzero Lyapunov
exponents, however, a difference between the perturbed model ODE and the high-symmetric flow
simulations on one side, and the unperturbed model ODE on the other side, 
becomes apparent. In the perturbed model ODE and in the simulations
the two largest nonzero exponents remain separate
in between doublings, or at least no crossing is found with a small but finite step size
in the parameter. An explanation might follow from a detailed study of the model ODE (\ref{normform})
with nonzero perturbation. Other interesting questions would be 
if the scaling laws of the doubling cascade
are influenced by the perturbation and how many steps of the cascade can be observed given
the ratio of fundamental frequencies and the strength of the perturbation.

The solutions presented here, in particular the bifurcating tori, 
are solutions of the truncated three dimensional Navier-Stokes
equations on a periodic domain. They might, however, prove to be unstable to asymmetric
perturbations. It is therefore uncertain whether the quasi-periodic doubling cascade
can be observed in periodic flow without any symmetries imposed. It does, however, occur at
low Reynolds number, at which asymmetric perturbations might be damped. Even at such low
Reynolds number, the simulation of general three dimensional flow, and the bifurcation
analysis as presented here, remains a formidable task. 

\ack
I would like to thank Marco Martens, who suggested the model ODE construction in section
\ref{qpdc}, Shigeo Kida, Henk Broer, Yuri Kuznetsov and Carles Sim\'o for useful discussions and
Greg Lewis for his hospitality at the University of Ontario Institute of Technology. This work was
supported by the Japan Society for Promotion of Science.

\appendix

\section{An explicit suspension of the H\'enon map}
\label{suspension}

Explicit expressions for a vector field that yields a fixed time, smooth suspension of the H\'enon
map were given by Mayer-Kress and Haken \cite{mayer}. Here we repeat the essential definitions in
a the notation of the model ODE \ref{normform}.

The H\'enon map is given by
\begin{eqnarray}
\mathcal{H}_{s}\left(\begin{array}{c} x \\ y \end{array}\right) &=& \left(\begin{array}{c} y+F(x;s) \\ -jx \end{array}\right)
\nonumber \\
F(x;s) &=& 1-sx^2
\end{eqnarray}
In order to define the function $\mathbf{f}(x,y,u;s)$ we need to introduce the following auxiliary functions:
\begin{eqnarray}
\xi(u) &=& (3-2u)u^2 \nonumber \\
\eta(u) &=& 1+ u\ln j -(2\ln j +3)u^2 +(2+\ln j)u^3 \nonumber \\
G(\xi,\eta) &=& \frac{1}{\xi^2+\eta^2}\left\{ \xi'\eta-\xi\eta'-\xi\eta\left(\ln j -2
\frac{\xi\xi'+\eta\eta'}{\xi^2+\eta^2}\right)\right\} \nonumber \\
H(\xi,\eta) &=& \frac{1}{\xi^2+\eta^2}\left\{(\xi'\xi+\eta'\eta)\frac{\xi^2-\eta^2}{\xi^2+\eta^2}
+\eta^2 \ln j\right\} \nonumber \\
X(x,y) &=& j^{-u}(\eta x -\xi y)
\end{eqnarray}
where the primes denote differentiation with respect to $u$. Suppressing the explicit dependence
on $u$ in the right hand side we have
\begin{equation}
\mathbf{f}(x,y,u;s)=\left( \begin{array}{cc} H(\xi,\eta) & G(\xi,\eta) \\
G(\eta,\xi) & H(\eta,\xi) \end{array} \right) \left(\begin{array}{c} x \\ y \end{array}\right) +
F(X(x,y);s)\left(\begin{array}{c} \xi\xi' \\ \xi'\eta \end{array}\right)
\end{equation}


\begin{thebibliography}{10}
\bibliographystyle{elsart-num.bst}

\bibitem{brachet} Brachet M E, Meiron D I, Orszag S A, Nickel B G, Morf R H and Frisch U 1983
Small-scale structure of the Taylor-Green vortex
{\it J.Fluid Mech.} {\bf 130} 411-452.

\bibitem{kida1}
Kida S 1985 Three-dimensional periodic flows with high-symmetry 
{\it J. Phys. Soc. Japan} {\bf 54} 2132--2136.

\bibitem{kida3}
Kida S and Murakami Y 1989 Statistics of velocity gradients in turbulence at moderate Reynolds number
{\it Fluid.Dyn.Res.} {\bf 4} 347--370.

\bibitem{kida4}
Kida S and Murakami Y 1987 Kolmogorov similarity in freely decaying turbulence
{\it Phys.Fluids} {\bf 30} 2030--2039.

\bibitem{boratav}
Boratav O N and Pelz B P 1997 Structures and structure functions in the inertial
range of turbulence
{\it Phys. Fluids} {\bf 9} 1400--1415.

\bibitem{lust}
Lust K, Roose D, Spence A and Champneys A R 1998 An adaptive Newton-Picard algorithm with
subspace iteration for computing periodic solutions
{\it SIAM J.Sci.Comput.} {\bf 19} 1188-1209.

\bibitem{sanchez}
S\'anchez J, Net M, Garc\'\i a-Archilla B and Sim\'o C. 2003 Newton-Krylov continuation of periodic
orbits for Navier-Stokes flows
{\it J.Comput.Phys.} in press.

\bibitem{kida2}
Kida S, Yamada M and Ohkitani K 1989 A route to chaos and turbulence
{\it Physica D} {\bf 37} 116--125.

\bibitem{kaneko}
Kaneko K 1986 {\it Collapse of tori and genesis of chaos in dissipative systems}, World Scientific,
Singapore.

\bibitem{arneodo}
Arn\'eodo A, Coullet P H and Spiegel E A 1983 Cascade of period doublings of tori
{\it Phys. Lett. A} {\bf 94} 1--5.

\bibitem{simo}
Sim\'o C 1989 On the analytical and numerical approximation of invariant manifolds
{\it Modern Methods in Celestial Mechanics, Proceedings of the 13th spring school 
on astrophysics in Goutelas} Edition Fronti\`eres, Gift-sur Yvette, France.
Available via {\tt http://www.maia.ub.es/dsg/2004/index.html}

\bibitem{broer}
Broer H W, Huitema G B and Takens F. 1990 Unfoldings of quasi-periodic tori
{\it Mem. AMS} {\bf 83}(421) 1--82.

\bibitem{braaksma}
Braaksma B L J, Broer H W and Huitema G B 1990 Toward a quasi-periodic bifurcation theory
{Mem. AMS} {\bf 83}(421) 83--175.

\bibitem{schilder}
Schilder F, Osinga H M and Vogt W 2004 Continuation of quasi-periodic invariant tori
{\it Applied nonlinear dynamics preprint} 2004.11 University of Bristol.

\bibitem{anishenko}
Anishenko V S 1990 {\it Complex oscillations in simple systems}, Nauka, Moscow.

\bibitem{zhong}
Zhong G-Q, Chai W W and Chua L O 1998 Torus-doubling bifurcations in four mutually coupled
Chua's circuits
{\it IEEE Trans. Circuits Syst. I} {\bf 45} 186--193.

\bibitem{franceschini}
Franceschini V 1983 Bifurcation of tori and phase locking in a dissipative system of differential equations
{\it Phys. D} {\bf 6} 285--304.

\bibitem{broer2} 
Broer H W, Sim\'o C and Vitolo R 2002 Bifurcations and strange attractors in the Lorenz-84 climate 
model with seasonal forcing
{\it Nonlinearity} {\bf 15} 1205--1267.

\bibitem{stag}
Stagliano J J, Wersinger J-M and Slaminka E E 1996 Doubling bifurcations of destroyed $\mathbb{T}^2$ tori
{\it Phys. D} {\bf 92} 164--177.

\bibitem{mayer}
Mayer-Kress G and Haken H 1987 An explicit construction of a class of suspensions and autonomous
differential equations for diffeomorphisms in the plane
{\it Commun.Math.Phys.} {\bf 111} 63--74.

\bibitem{feig}
Feigenbaum M J 1980 The transition to aperiodic behaviour in turbulent systems
{\it Commun.Math.Phys.} {\bf 77} 65--86.

\end{thebibliography}
\end{document}